\documentclass[notitlepage,twocolumn,prl]{revtex4-1}
\usepackage{amsmath}
\usepackage{mhchem} 
\usepackage{changes}
\usepackage{graphicx}
\usepackage[space]{grffile} 
\usepackage{gensymb}
\usepackage{bm}
\usepackage{amssymb,amsfonts,amsmath,amsbsy,bm,t1enc,lipsum,latexsym}

\begin{document}

\title{Photonic chip-based soliton frequency combs covering the biological imaging window}

\author{Maxim Karpov}
\author{Martin H. P. Pfeiffer}
\author{Tobias J. Kippenberg}
\email{tobias.kippenberg@epfl.ch}
\affiliation{{\'E}cole Polytechnique F{\'e}d{\'e}rale de Lausanne (EPFL), Laboratory of Photonics and Quantum Measurements (LPQM), Lausanne, CH-1015, Switzerland}

\date{\today}

\begin{abstract}
Dissipative Kerr solitons (DKS) in optical microresonators provide a highly miniaturized, chip-scale frequency comb source with unprecedentedly high repetition rates and spectral bandwidth. To date, such soliton frequency comb sources have been successfully applied in the optical telecommunication band for dual comb spectroscopy, coherent telecommunications,  counting of optical frequencies and distance measurements. Yet the range of applications could be significantly extended by operating in the near-infrared spectral domain, a prerequisite for biomedical and Raman imaging applications, and a part of the spectrum which hosts commonly used optical atomic transitions. Here we demonstrate the operation of photonic chip-based soliton Kerr combs pumped with 1 micron laser light. By engineering the dispersion properties of a $\rm Si_3N_4$ microring resonator, octave-spanning soliton Kerr combs extending to 776 nm are attained,  thereby covering the optical biological imaging window. Moreover, we demonstrate that soliton states can be generated in normal group velocity dispersion regions, when exploiting mode hybridization with other mode families. The reported near-infrared soliton Kerr combs are suitable for a range of studies, including frequency-comb-based optical coherence tomography, coherent anti-Stokes Raman spectro-imaging, and they spectrally overlap with optical atomic transitions in Alkali vapors.
\end{abstract}

\maketitle

\noindent \textbf{Introduction.} ---
Microresonator-based frequency combs are optical frequency combs generated from a continuous-wave (CW) laser via parametric four-wave mixing processes in high-\emph{Q} microresonators \cite{delhaye2007comb}. They have attracted significant interest owing to their compactness, wafer-scale integration, and ability to operate with repetition rates in the microwave to terahertz range with broad bandwidth \cite{Kippenberg2011microcombs}.
It was recently demonstrated that such CW laser-driven microresonators can support the spontaneous formation of dissipative Kerr solitons \cite{herr2014soliton} (DKS) -- self-organized stable intracavity pulses relying on the double balance between dispersion and nonlinearity \cite{Akhmediev2003}, and parametric gain and cavity losses  -- which provide a route to fully coherent optical frequency combs whose spectral bandwidth can be significantly broadened via soliton-induced Cherenkov radiation \cite{Brasch2014Cherenkov, Li2017octaveDKS, pfeiffer2017octave}. Such microresonator-based DKS combs were demonstrated in several platforms \cite{herr2014soliton, Brasch2014Cherenkov,Yi2015solitoncaltech,webb2016DKSmicrospheres}, and have already been successfully applied in optical coherent communications \cite{marin2016DKScommunication}, dual-comb spectroscopy \cite{dualcomb2016vahala,yu2016DKSMIR}, implementation of the microwave-to-optical link via self-referencing \cite{brasch2017DKSselfreferencing, Jost2014link}, and most recently in dual-comb distance measurements \cite{Ganin2017distancemeasurements,suh2017distancemeasurements}.
Yet, to date, well-identified and single DKS have only been attained in the telecommunication bands (around 1550 nm), where a wide variety of DKS-supporting microresonator platforms have been developed, and more recently DKS have been extended into the mid-infrared (MIR) \cite{yu2016DKSMIR}.

Importantly, a large class of new applications of biomedical nature can be accessed with DKS-based comb sources if they can operate in the short-wavelength part of the near-infrared (NIR) domain ranging from 0.7 to 1.4 $\mu$m wavelength. This spectral region is used for biological and medical imaging techniques due to the highest penetration depth in biological tissues limited by the water and blood absorption outside of this spectral window \cite{smith2009NIRwindowbio}. Optical spectroscopy, Raman spectro-imaging and optical coherence tomography (OCT) techniques operating in this wavelength range serve as a non-invasive tool for the structural and chemical analysis of the various biological samples, including retinal and choroidal structures or tumor formations \cite{drexler2001ophtalmicOCT,reich2005NIRspectroscopy,richards1996quantOptAbsorption}.  These biomedical imaging techniques could benefit from employing optical frequency combs as sources due to their coherence and high power per comb line \cite{Siddiqui2017combOCT}, as well as from using dual-comb approaches allowing to interrogate the broad spectral bands with a single photodetector \cite{coddington2016dualcomb}. A specific example of dual-comb-based spectroscopy in the NIR is coherent anti-Stokes Raman spectroscopy (CARS) \cite{ideguchi2013coherent}, which can utilize high repetition rates of the DKS combs for vastly increased acquisition rate, enabling real-time CARS imaging. Furthermore, luminescent-free anti-Stokes response can benefit from the larger Raman cross section and reduced focal spot size at shorter wavelength, facilitating phase matching.  Such dual-DKS-comb CARS may be able to provide ultrafast multispectral \emph{in-vivo} imaging for chemical, biological and medical purposes. Equally important, a range of other applications requiring the stabilized operation can be also accessed and improved by near-infrared DKS-based combs. The NIR domain hosts optical frequency standards in Alkali vapours (e.g.\ce{^{87}Rb}, \ce{^{133}Cs}), necessary to realize chip-scale optical atomic clocks with enhanced precision \cite{Papp2014kerropticalclock}, or employed in DKS-comb-based calibration for astronomical spectrometers \cite{quinlan2012astrocomb}.

\begin{figure*}
\includegraphics[width = \textwidth]{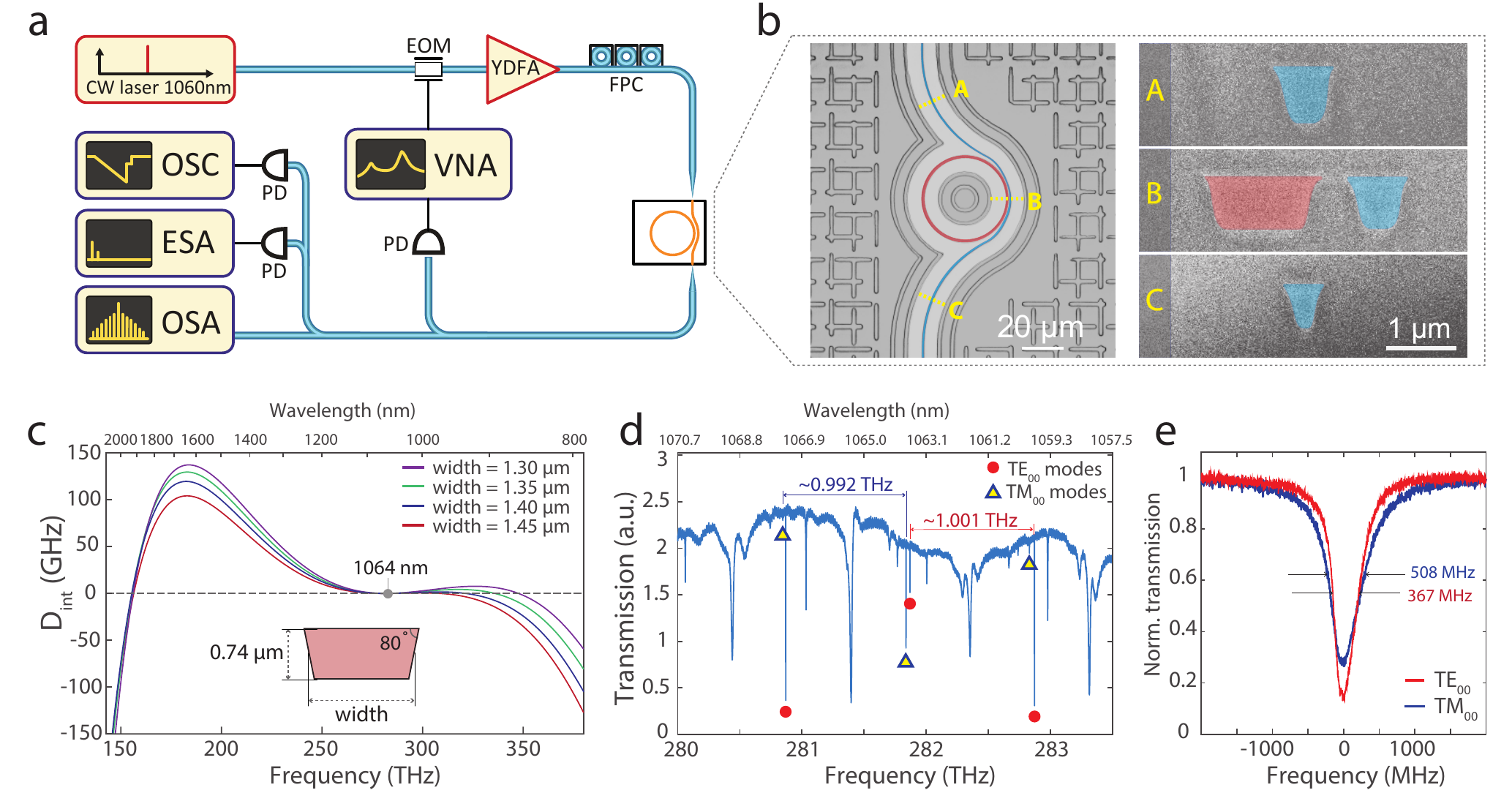}
\protect\caption{\textbf{Design and characterization of \boldmath{$\rm Si_3N_4$} microresonator for soliton frequency comb generation in the NIR}
(a) Set-up scheme used for DKS generation and characterization: A tunable external-cavity diode laser with a center wavelength of 1050 nm is used as a seed, YDFA -- ytterbium-doped fiber amplifier, FPC -- fiber polarization controller, VNA -- vector network analyzer, EOM -- electro-optical phase  modulator, PD -- photodioide, OSC -- oscilloscope, ESA -- electrical spectrum analyzer, OSA -- optical spectrum analyzer. 
(b) Left: Optical microscope image of the 1-THz microring resonator (highlighted in red) with a pulley-style bus waveguide (blue). Right: Scanning electron microscope images of the resonator and bus waveguide cross sections obtained via focused ion beam at different positions (A-C) marked on the left image.
(c) Simulated integrated dispersion profiles ($D_{\rm int}$) for $\rm TM_{00}$ mode of resonator waveguides having common height of 0.74 $\mu$m, sidewall angle of 80\degree, and different widths from 1.3 to 1.45 $\mu$m (see inset for the geometry details). 
(d) Transmission trace of the 1-THz microresonator shown in Fig.1 (b). The two fundamental mode families ($\rm TE_{00}$ and $\rm TM_{00}$) can be  distinguished based on their FSR (0.992 THz for $\rm TM_{00}$ and 1.001 THz for $\rm TE_{00}$), and are marked with red and blue shapes correspondingly. Other resonances correspond to higher order modes with comparably lower \emph{Q}-factors. 
(e) Linewidth measurements of the fundamental mode families. The frequency axis was calibrated using a fiber-loop cavity. Typical loaded linewidth of the modes is $\sim400$ MHz for $\rm TE_{00}$ and $\sim500$ MHz for $\rm TM_{00}$. 
\label{fig_1}}
\end{figure*}

Despite this large number of promising applications of the near-infrared DKS-based combs, such sources have so far not been developed. 
Although attempts to generate Kerr combs in NIR and visible domain have been made \cite{saha2012Kerrcomb1um,guo2017visKerrcombSHG,wang2016greencomb,savchenkov2011visKerrcomb,yang2016FWMvis}, they resulted in relatively narrow  and incoherent combs, and soliton formation has not been achieved.
The generation of the NIR or visible soliton combs is compounded by the increased normal group velocity dispersion (GVD) of the materials due to the bandgap electronic absorption in UV, the increased scattering losses and sensitivity to the resonator waveguide dimensions, which require higher precision of dispersion engineering and fabrication processes. Moreover, as shown recently, competition between Raman and Kerr effects in the NIR or visible domains can inhibit soliton formation \cite{okawachi2017ramanCompetition}.

\begin{figure*}
\includegraphics[width = \textwidth]{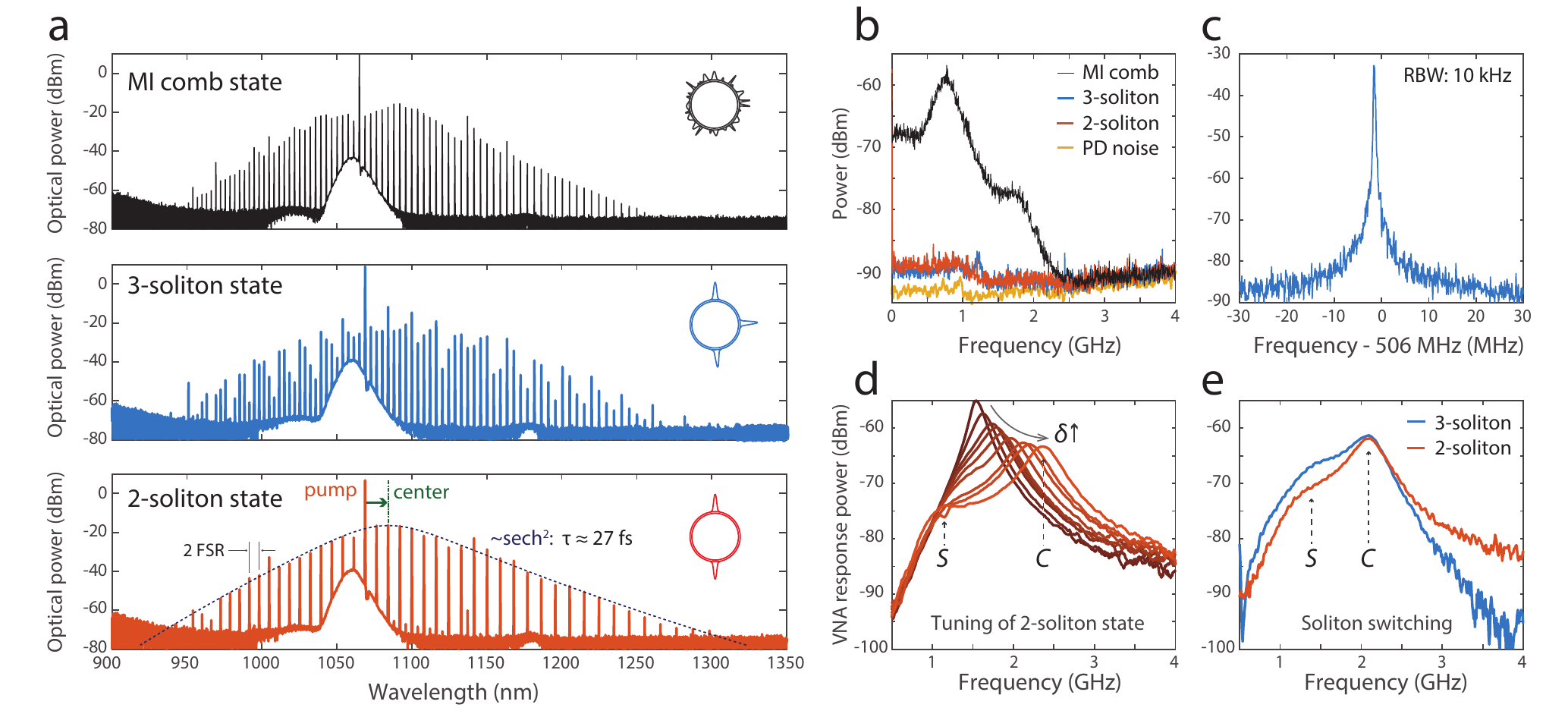}
\protect\caption{\textbf{Dissipative Kerr solitons at 1 $\mu$m and their characterization}
(a) Optical spectra of the MI comb state (top) and two soliton states (middle, bottom) obtained in a 1-THz $\rm Si_3N_4$ microresonator. The transition between 3-soliton state (middle) to 2-soliton state (bottom) was obtained by a backward tuning technique of the pump laser. Insets show the estimated positions of the DKS in corresponding states. The 2-soliton state was fitted with a $\rm sech^2$ envelope (dashed dark blue), with estimated duration of 27 fs. The green arrow shows the Raman-induced red spectral shift of the soliton spectrum with respect to the pump line.
(b) Amplitude noise of MI comb state (black) and soliton states (blue, red) shown in Fig.1(a). 
(c) Measured beat note between the comb line at 1050 nm  of the Dependence of the $\mathcal{S}$-resonance with the pump power, measured for a stationary soliton. This resonance provides an estimate of the detuning point at (d)
(e)
\label{fig_2}}
\end{figure*} 

In this work, we demonstrate that microresonators can overcome these challenges and DKS-based frequency combs can be generated with a 1064-nm CW laser, thereby allowing access to the optical wavelength window for biological imaging (0.7 -- 1.4 $\mu$m). The typical signatures \cite{herr2014soliton,karpov2015raman,karpov2016universal} of dissipative Kerr solitons in microresonators, including the low-phase-noise operation, Raman-induced red spectral shift, soliton switching dynamics and the bistability-related double-peak response of the microresonator system are observed and provide unambiguous identification of the DKS states.  Equally important, we demonstrate the formation of octave-spanning soliton states in this wavelength region by exploiting the coherent dispersive wave emission for efficient spectral broadening towards visible wavelengths. Finally, we report on soliton formation in hybridized microresonator modes, which can represent an alternative approach to extend the DKS operation further into the visible domain, where normal GVD is usually dominant. The DKS states are achieved in the region of avoided modal crossing, where the strong interaction between resonator modes leads to locally enhanced anomalous GVD, allowing the DKS formation. We show that the bandwidth of such soliton states is highly sensitive to pumped resonance within the interaction region -- a behaviour contrasting the behavior in the absence of modal crossings, and a behavior not observed before.

\noindent \textbf{Device design and characterization} --- 
\label{Device design and characterization}
We employed the silicon nitride ($\rm Si_3N_4$) microresonator platform, which is a well-developed and extensively used basis for on-chip nonlinear and quantum photonics due to a number of advantages such as CMOS-compatibility, high effective nonlinearity, negligible two-photon absorption, and wide transparency window spanning from visible to mid-infrared \cite{Levy2010SINcomb,li2016efficient,moss2013cmos}.  Recent advances in the fabrication processes have enabled the fabrication of crack-free low-loss $\rm Si_3N_4$ waveguides with void-free coupling gaps, which guarantee high-\emph{Q} resonators with well-controllable properties \cite{gondarenko2009high,Pfeiffer2015Damascene,herkommer2017midirdispwave,pfeiffer2017octave,kordts2015higher}. An important advantage of that is the ability to engineer the dispersion properties of the microresonators, by compensating the material dispersion with the waveguide dispersion contribution. \cite{Brasch2014Cherenkov,Li2017octaveDKS,pfeiffer2017octave}.
In the context of microresonator-based Kerr frequency combs, the dispersion properties are often expressed through the frequency deviations from an equidistant grid for a certain pumped mode $\omega_0$ in the relative mode index (${\mu}$) representation: ${D_{\rm int}(\mu) = \omega _{\mu } - (\omega_0 + D_{1} \mu ) = \sum_{i>1} D_{i}\mu^{i}/{i!}}$, ${i\in \mathbb{N}}$, where  ${\omega_{\mu}}$ is the angular frequency of the cavity resonance, and $\frac{D_{1}}{2\pi}$ is the microresonator free spectral range (FSR) \cite{herr2014soliton}. For bright DKS generation it is generally required to achieve anomalous group velocity dispersion (GVD): $D_{2} >0$, which can be especially challenging at short wavelengths, due to the presence of the material's UV absorption.

In this work we used $\rm Si_3N_4$ microrings with FSR of $\sim $1 THz (radius $\sim$23 $\mu$m, see Fig.1 (b)), which were fabricated using the \emph{Photonic Damascene process} \cite{Pfeiffer2015Damascene}. The resonator waveguide width was varied from 1.3 to 1.5 $\mu$m, and the height was targeted at 0.74 $\mu$m (with process-related variations on the order of 10 nm), which were chosen based on FEM simulations (see Fig.1(c)) in order to ensure anomalous GVD of the fundamental TM mode around the pumping wavelength of 1060 nm. The bus waveguide has pulley-style coupling section with altering width from $0.3~\rm \mu m$ at the in-coupling part (see Fig.1(b) - C) to $0.65~\rm \mu m$ at the out-coupling part (see Fig.1(b) - A,B), which was designed to guarantee broadband coupling, with however limited ideality \cite{pfeiffer2017ideality}. The waveguides have inverse tapered structures (down to $\sim$150 nm width) at the input and output ends, providing $< 3$ dB coupling loss per facet at 1$\mu$m.

A typical transmission trace of the fabricated devices is shown in Fig.1(d). The comparably large width of the microring waveguide induces significant overmoding of the resonator. Two fundamental modes  -- $\rm TE_{00}$, $\rm TM_{00}$ -- can be easily identified due to different FSRs (0.992 THz for $\rm TM_{00}$ and 1.001 THz for $\rm TE_{00}$) provided by the non-unity aspect-ratio, and are slightly undercoupled for the available gap distances. The modes have comparable loaded linewidths of 370 - 550 MHz, corresponding to the \emph{Q}-factor of $0.55-0.75\times10^6$.

\begin{figure*}
\includegraphics[width = \textwidth]{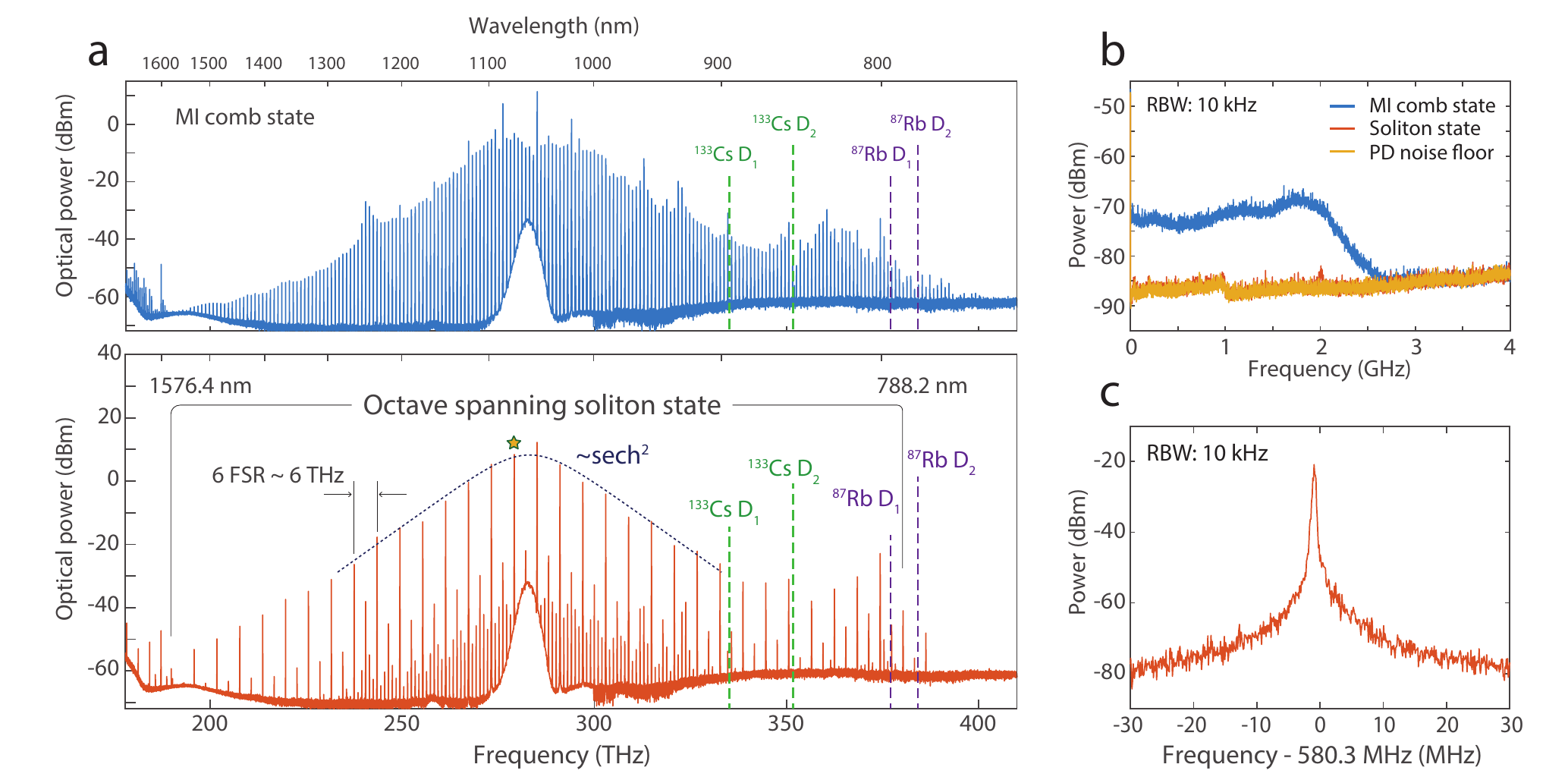}
\protect\caption{\textbf{Octave-spanning microresonator-based dissipative Kerr soliton state in the biological imaging window}
(a) Top: Spectrum of a noisy Kerr comb state, obtained with $\sim$800 mW on-chip power (pump is located at around 1051.5 nm), from the microresonator with optimized dispersion for octave-spanning operation. Green and purple dashed vertical lines indicate the spectral locations of Cs and Rb optical atomic transitions. Bottom: Spectrum of the octave-spanning soliton state obtained in the same microresonator. The spectrum is fitted with $\rm sech^2$ envelope (dashed dark blue), from which a soliton duration of 17 fs is inferred. The set of enhanced lines below 190 THz corresponds to the second diffraction order of the OSA's diffraction grating and are thus artefacts. 
(b) Intensity noise of the octave-spanning MI comb state (blue) and soliton state (red) shown in (a). Yellow trace shows the noise floor of the photodetector used for the measurements.  
(c) Heterodyne beatnote of the soliton comb line (marked with a star) with a second CW laser.
\label{fig_3}}
\end{figure*} 

\noindent \textbf{Dissipative Kerr solitons at 1$\mu$m} ---
\label{DKS at 1um}
The experimental set-up for the DKS generation in $\rm Si_3N_4$ microresonators is shown in Fig.1(a). The light from a tunable 1050-nm CW diode laser (Toptica CTL 1050) is amplified with an ytterbium-doped fiber amplifier and coupled to the chip through a lensed fiber. The output signal was collected with another lensed fiber, and it's spectral and noise characteristics were analyzed. We also employ a recently developed soliton probing technique \cite{karpov2016universal}, which uses a phase-modulated pump and a vector network analyser to retrieve a system response allowing us to unambiguously identify DKS formation and track the detuning of the pump from a cavity resonance. 

It is well-known, that the CW-driven nonlinear cavity can support DKS states, when operating in the effectively red-detuned regime ($\omega_p-\omega_0=2\pi \delta > 0$, where $\omega_p$ and $\omega_0$ are the angular frequencies of the pump laser and pumped resonance). A standard approach for accessing such states is the laser tuning method \cite{herr2014soliton,karpov2016universal}, where the pump laser is swept over a resonance from the blue to the red side, and is stopped at a certain pump-resonance detuning ($2\pi\delta > \frac{\sqrt{3}}{2}\kappa$) supporting soliton formation. For the samples used in our work it is sufficient to apply this standard approach, rather than complex techniques such as power kicking \cite{Brasch2014Cherenkov} or fast frequency modulation \cite{Stone2017SSBsoliton}.

We apply this technique with an on-chip pump power of $\sim1~$W to the resonances of the fundamental TM mode family, which should provide anomalous GVD at around 1060 nm. Low tuning speeds on the order of a few GHz/second (i.e.using laser cavity piezo) were chosen and enabled to simultaneously monitor the cavity state by measuring the system response signal as well as an optical spectrum of the output light. The cavity reveals modulation instability and noisy Kerr comb formation while the pump laser is on a blue side of the resonance (see Fig.2(a, top)). Upon further pump tuning, the transition to the soliton regime is accompanied by a change of the optical spectrum to the secant hyperbolic-like shape, and an appearance of the double-peak structure in the system response representing the coexistence of the soliton ($\mathcal{S}$-resonance) and CW-background ($\mathcal{C}$-resonance) components inside the cavity. To explore the soliton existing range of the generated DKS state, the response measurements were also carried out, while tuning the pump laser towards longer wavelength corresponding to the increase of the effective detuning (Fig.2(c)). The expected shift of the position of the $\mathcal{C}$-resonance (which has been shown to indicate the effective detuning of the system) to higher frequencies is clearly seen and essentially reproduces the dynamics of similar response signals measured for DKS at 1550 nm \cite{karpov2016universal}. Finally, the transition to the soliton regime has been also verified by the drastic reduction of the output light intensity noise (see Fig.2(b)), and a narrow heterodyne beatnote of the selected comb line at around 1050 nm with another CW diode laser (Fig.2 (c)).

We also demonstrate that the obtained soliton states can experience switching by applying the recently-reported backward tuning technique, which relies on the thermal nonlinearity of microresonators, and allows the number of DKS circulating inside the cavity to be changed in a robust and controllable way \cite{karpov2016universal}. Figure 2(a) shows the switching from a 3- to 2-soliton state, where the final state consists of two solitons separated by almost 180 degrees, which has resulted in the two-FSR spacing of the optical spectrum. The switching has been also confirmed with the response measurements as shown in figure 2(e). A decrease in the amplitude of the soliton-number-related $\mathcal{S}$-resonance of the response indicates the reduction of the intracavity number of pulses, while the cavity-related $\mathcal{C}$-resonance is almost unchanged. By fitting the later DKS spectrum with a $\rm sech^2$ envelope, the soliton duration can be estimated from its 3-dB bandwidth as 27 fs. We also note the significant soliton red spectral shift ($\sim 4.1$ THz in the present case) with respect to the pump line, which is mainly attributed to the Raman effect and observed for all DKS states above \cite{milian2015interplayRamanTOD,karpov2015raman}.

\noindent \textbf{Octave-spanning soliton states} ---
\label{Octave-spanning}
Reaching the octave-spanning operation of DKS is a useful step in the development of Kerr frequency combs, as it enables the common $f-2f$ scheme for offset frequency detection and self-referencing required by multiple applications in optical frequency metrology and low-noise microwave synthesis \cite{newbury2011searchapplications}. Octave-spanning DKS states have been only very recently demonstrated experimentally \cite{Li2017octaveDKS,pfeiffer2017octave}. 
Here we demonstrate that DKS-based combs operating at 1$\mu$m can be engineered to have octave-spanning bandwidth despite operating close to the normal GVD region in the $\rm Si_3N_4$ platform.

We used a microresonator with the same FSR of 1 THz as in the previous section, and having a waveguide geometry of $1.30 \times 0.74~\mu$m, which was designed to maintain low anomalous GVD and satisfy phase-matching conditions at around 800 nm for dispersive wave formation \cite{Brasch2014Cherenkov}. This spectral region is particularly interesting due to the presence of optical frequency standards based on the two-photon Rb transitions \cite{jones2000fceostabilization} which can be used for the comb referencing. We applied the aforementioned low-speed tuning technique with an estimated on-chip power of $\sim$800 mW to achieve the formation of the noisy comb (see Fig.3(a, top)) followed by the soliton state (Fig.3(a, bottom)). 

The resulting spectrum of the soliton state spans over an octave from 776 to 1630 nm ( $>200$ THz). As expected, it is significantly extended towards shorter wavelengths due to the emission of the dispersive wave via soliton-induced Cherenkov radiation at 800 nm \cite{akhmediev1995cherenkov}. The 3-dB bandwidth of the spectrum fitted with a $\rm sech^2$ envelope is estimated as 18 THz, which corresponds to the $\sim$18-fs pulses. A peculiar shape of the DKS state, consisting of several soliton-like spectra with different FSR is attributed to the formation of a multiple-soliton state (which here is estimated to be a six-soliton state) represented by the ordered co-propagating DKS ensemble -- soliton crystal \cite{cole2016solitoncrystal,Karpov2017solitonCrystal}. Such soliton crystals are typically formed in the presence of strong local spectrum deviations caused by inter-mode interactions among transverse mode families (avoided mode crossings), and in contrast to single-soliton states are featuring high conversion efficiency (owing to the high number of intracavity pulses), which in the present case approached 50\%.

Similarly to the DKS states demonstrated in the previous section, the presented state is also characterized by a low-noise perfomance with a strongly suppressed intensity noise in comparison to the noisy MI comb (see Fig.3(b)), and a narrow heterodyne beatnote of the generated comb line with another CW laser (Fig.3(c)).

\begin{figure*}
\includegraphics[width = \textwidth]{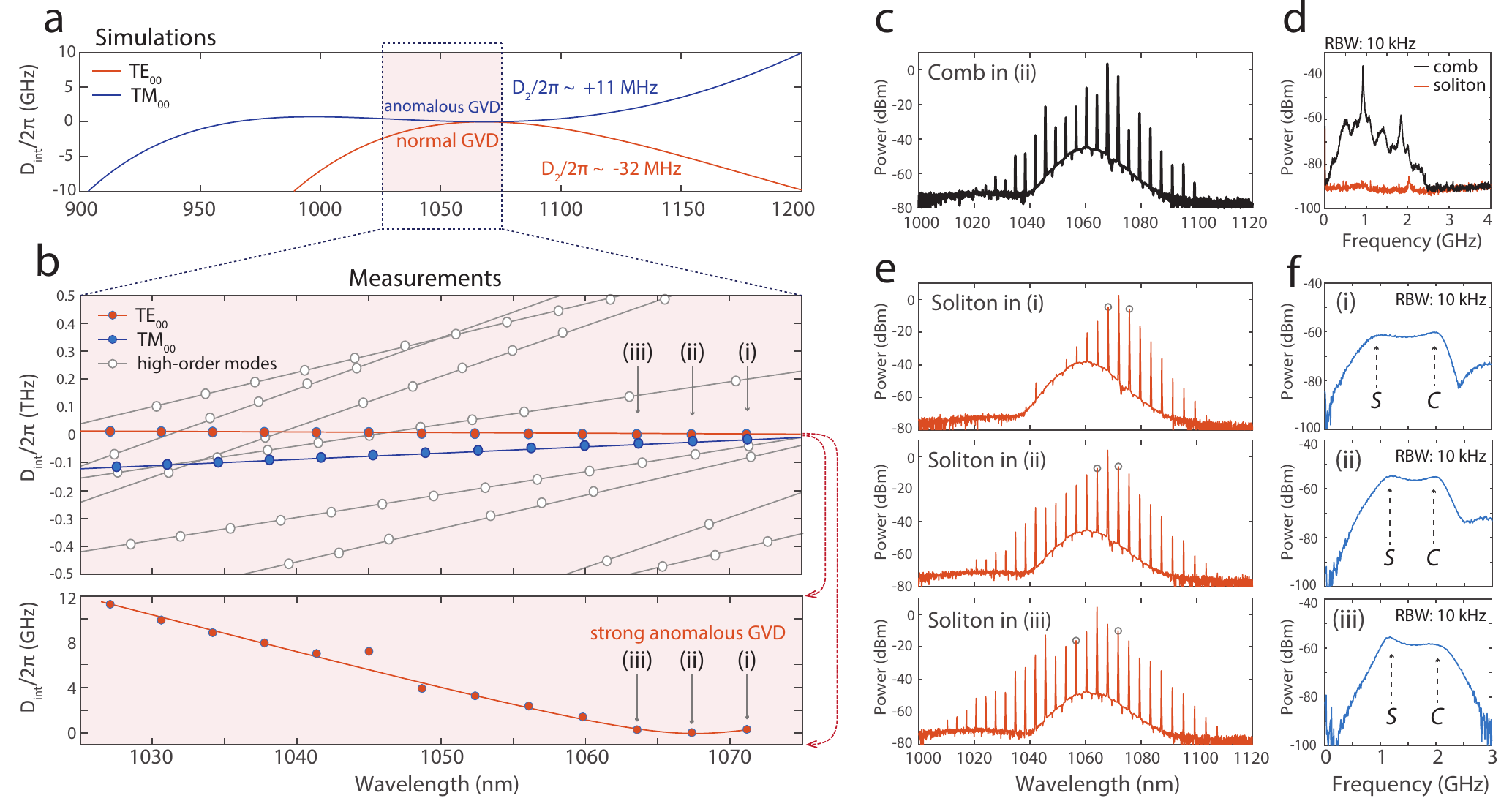}
\protect\caption{\textbf{Dissipative Kerr solitons in hybridized modes}
(a) Simulations of the integrated dispersion of the fundamental TE and TM modes in the microring resonator with a waveguide dimensions of $1.45 \times 0.74~\mu$m. The $\rm TE_{00}$ mode has normal GVD with $\frac{D_{2}}{2\pi} = -32$ MHz, while the $\rm TM_{00}$ mode has anomalous GVD with $\frac{D_{2}}{2\pi} = 11$ MHz.
(b) Frequency-comb assisted measurements \cite{Delhaye2009disp} of the mode structure of a 1-THz $\rm Si_3N_4$ microring resonator with the dimensions of $1.45 \times 0.74~\mu$m simulated in (a). Mode families are distinguished based on their FSRs. General dispersion trend cannot be identified due to the bandwidth limitations of our measurement setup, however a strong local dispersion for three consecutive modes (i, ii, iii) of the $\rm TE_{00}$ mode family above 1064 nm can be observed.
(c) Optical spectrum of a noisy comb state, obtained in the resonance ii  from (b).
(d) Intensity noise measurements of noisy comb (comb, black) and the DKS (soliton, red) states, obtained by pumping the resonance (i) from (b). 
(e) Optical spectra of the DKS states obtained by pumping resonances i, ii, iii from (b). Grey circles indicate the positions of primary comb lines.
(f) Response measurements of DKS states represented in (e). The positions of characteristic $\mathcal{C}$- and $\mathcal{S}$-resonance are indicated with \emph{C} and \emph{S} letters correspondingly.
\label{fig_4}}
\end{figure*}

\noindent \textbf{Dissipative Kerr soliton states in hybridized modes} ---
\label{Hybridized modes}
The behavior of Kerr combs, and in particular DKS states are in large extent defined by the dispersion properties of the cavity. Apart from the global dispersion landscape comprised of the material- and waveguide-related components, such properties also include spectrally localized dispersion modifications (avoided mode crossings, AMX) that are typically caused by the formation of guided hybridized modes, which can appear due to the interaction of different transverse mode families. It was recently shown that such AMX-s can lead to complex and diverse effects on the dynamics of DKS states, such as dispersive wave formation, soliton recoil, temporal soliton ordering, appearance of quiet operation points and inter-mode soliton breathing. \cite{herr2014soliton,guo2017inter-modeBreathers, yi2017SMdispersiveWave, cole2016solitoncrystal, Karpov2017solitonCrystal}. Moreover, AMX has been reported to allow the formation of mode-locked states consisting of dark pulses in resonators with normal GVD \cite{xue2015darksol}.

In this section we demonstrate that localized strong anomalous GVD of the \emph{hybridized} modes around AMX can be directly employed for bright soliton generation \emph{irrespective} of the global dispersion profile. Figure 4(b) shows the measured \cite{Delhaye2009disp} mode structure of one of the fabricated samples with a 1.45 $\mu$m waveguide width, where the corresponding modes of the fundamental TE and TM mode families are highlighted. Extrapolating the mode families to longer wavelength beyond the measurement range (limited with the tunability of our laser), an interaction region of multiple transverse modes at around 1080 nm is expected. Here we particularly focus on the TE mode family, which according to our simulations has normal GVD (see Fig.4(a)). A closer look at its integrated dispersion reveals significant deviation from the uniform trend of several consecutive resonances above 1060 nm, caused by the interaction of multiple mode families (see Fig.4(b), bottom). The strong resonance shifts cause a dramatic change on the local GVD of the TE mode family turning it from normal with $\frac{D_{2}}{2\pi} = -32$ MHz (simulated) to highly anomalous, with $\frac{D_{2}}{2\pi}$ reaching 510 MHz (measured). We therefore evidence an AMX-induced change in the local group velocity dispersion.

Driving the modes with such a strong anomalous GVD (when $\sqrt{\frac{\kappa}{D_{2}}} < 1 $) should result in the formation of natively mode-spaced comb \cite{ferdous2011linebylineshape,Herr2012universal}, whose primary lines appear 1-FSR away from the pump due to the closely-located MI gain peaks. Previous works have reported that such combs can appear directly in a mode-locked regime \cite{liu2014modecoupling}, which is however in contrast with our observations. In the experiments we again used the same pump tuning technique as in previous sections, applied to hybridized modes forming AMX. Using the system response and intensity noise measurements we track the standard soft-excitation-route of a Kerr comb formation, which includes the appearance of primary comb lines, development of the chaotic modulation instability (MI, Fig.4(c)) and the transition to a \emph{soliton} regime (Fig.4(e)), which was verified by the system response measurements (Fig.4(f)). Similar behaviour was observed over three consecutive resonances (see cases I, II, III in Fig.4(e)), where the corresponding soliton states were generated. An interesting observation can be made regarding the spectral bandwidth of the obtained soliton states. As a result of the localized character of the anomalous GVD of hybridized modes, the actual value of the dispersion terms (and particularly $\frac{D_{2}}{2\pi}$ term) can vary from one resonance to another, leading to the different effective detunings for generated comb lines and altering the resulting spectral width. We also note here, that due to the contribution of higher-order dispersion terms around AMX, the obtained DKS spectra cannot be faithfully fitted with a $\rm sech^2$ envelope.

The demonstrated soliton states generated in hybridized modes by exploiting their strong anomalous GVD can represent an alternative way to deterministically generate soliton-based optical combs in arbitrary wavelength regions. This approach can be especially useful in the platforms where the strong normal material GVD cannot be compensated with waveguide dispersion (as for example the $\rm Si_3N_4$ platform presented here for the visible wavelength) prohibiting the access to bright DKS.

\noindent \textbf{Discussion} ---
In conclusion, we show the first photonic-chip-integrated soliton-based optical frequency comb sources driven with a 1$\mu$m pump source. The spectra of the demonstrated DKS states are able to span over an octave, and cover the common optical frequency standards in Alkali vapors as well as significant part of the near-infrared window for biological tissues. Moreover, we show that DKS states can be generated in hybridized microresonator modes around avoided mode crossings by directly exploiting their localized anomalous GVD, which represents an alternative approach for the generation of DKS combs in the regions with strong normal GVD (e.g. at shorter wavelength in $\rm Si_3N_4$, and in other materials). From the broader perspective, our work gives strong evidence of the technological readiness of the $\rm Si_3N_4$ platform for soliton-based operation in the NIR domain around 1$\mu$m including comparably good quality factors and the means of dispersion engineering, which makes it a highly promising candidate for multiple biological and other applications in this spectral window, including OCT and dual-comb CARS.

\textit{Note from the authors} ---
We would like to draw the reader's attention to the other work \cite{lee2017visiblesoliton}, which was carried out concurrently with the present research and reports on the generation of DKS states centered at 1064 and 780 nm in fiber-coupled silica microdiscs.


\begin{acknowledgements}

\noindent \textbf{Acknowledgements} --- This publication was supported by Contract W31P4Q-14-C-0050 from the Defense Advanced Research Projects Agency (DARPA), Defense Sciences Office (DSO);
by the Air Force Office of Scientific Research, Air Force Material Command, USAF under Award No. FA9550-15-1-0099; 
We also acknowledge the support from  the European Space Technology Centre with ESA Contract No. 4000116145/16/NL/MH/GM and the support from the European Union's FP7 programme under under Marie Sklodowska-Curie Initial Training Network grant agreement No. 607493.
${\rm Si_3N_4}$ samples were fabricated and grown in the Center of MicroNanoTechnology (CMi) at EPFL.
\end{acknowledgements}


%

\end{document}